\documentclass[twocolumn,showpacs,amsmath,amstex,amssymb,mathfonts,prl]{revtex4-1}
\usepackage{amsthm,amsfonts,graphicx,verbatim,color}
\usepackage{amsmath}
\usepackage{amssymb}
\usepackage{amsthm}
\usepackage{amsfonts}
\usepackage{listings}
\usepackage{enumerate}
\usepackage{latexsym}
\usepackage{times}
\usepackage{psfrag}
\usepackage{bm}
\usepackage{graphicx}
\usepackage{subfigure}
\newcommand{\be}{\begin{equation}}
\newcommand{\ee}{\end{equation}}
\newcommand{\bea}{\begin{eqnarray}}
\newcommand{\eea}{\end{eqnarray}}

\renewcommand{\phi}{\varphi}
\renewcommand{\epsilon}{\varepsilon}

\begin{document}

\title{Quantum Phase Transitions and the $\nu=5/2$ Fractional Hall State in Wide Quantum Wells}

\author{Z. Papi\'c$^{1}$, F. D. M. Haldane$^1$, and E. H. Rezayi$^{2}$}
\affiliation{$^1$ Department of Physics, Princeton University, Princeton, NJ 08544} 
\affiliation{$^2$ Department of Physics, California State University, Los Angeles, California 90032, USA}


\date{\today}

\begin{abstract}
We study the nature of the $\nu=5/2$ quantum Hall state in wide quantum wells under the mixing of electronic subbands and Landau levels. A general method is introduced to analyze the Moore-Read Pfaffian state and its particle-hole conjugate, the anti-Pfaffian, under periodic boundary conditions in a ``quartered" Brillouin zone scheme containing both even and odd numbers of electrons.  By examining the rotational quantum numbers on the torus, we show spontaneous breaking of the particle-hole symmetry can be observed in finite-size systems. In the presence of electronic-subband and Landau-level mixing, the particle-hole symmetry is broken in such a way that the anti-Pfaffian is unambiguously favored, and becomes more robust in the vicinity of a transition to the compressible phase, in agreement with recent experiments.

\end{abstract}
\maketitle

The quantized Hall state at the $\nu=5/2$ Landau level (LL) filling factor~\cite{willett} has been the subject of 
significant recent interest due to a strong suspicion, with 
considerable support from the numerical calculations~\cite{morf,rh,storni,evenodd_lu}, that it is described by the Moore-Read 
``Pfaffian" (Pf) state~\cite{mr}. This incompressible quantum fluid,  which is a prototype state for non-Abelian
exchange statistics~\cite{mr,rg}, is found in the vicinity of the phase boundary with compressible
phases characterized by stripe order and Fermi-liquid-like behavior~\cite{rh}. Generally, the incompressible fluids of the fractional quantized Hall effect (FQHE)~\cite{tsg,laughlin,prange} possess
protected gapless edge modes, and gapped bulk excitations 
which carry fractional charge and obey fractional statistics~\cite{mr, zeromodes, statistics}. These attributes -- quantization, 
fractionalization, and protection -- represent the hallmark of topological phases~\cite{wen}. In the
case of non-Abelian states, the degeneracy of the quasi-particle states may be suitable to implement a ``fault-tolerant" quantum computation~\cite{tqc}.  

The Moore-Read state, though defined in a half-filled LL, is not invariant under the particle-hole (P-H) transformation, and therefore its P-H conjugate partner -- the ``anti-Pfaffian" (APf)~\cite{antipf}  -- emerged as a competing candidate to describe the ground state of $\nu=5/2$. 
In experiment, either Pf or APf is realized, depending on the explicit form of the P-H symmetry-breaking fields (e.g., 3-body interaction~\cite{donna} or more generally LL mixing~\cite{bn, llmixing1, llmixing2}). In the absence of those, the true ground state is selected by spontaneous  P-H symmetry breaking. A similar outcome should be reproducible in finite-size calculations, which have been known to capture remarkably well the fundamental aspects of FQHE physics~\cite{prange}.  However, for technical 
reasons (see below), this never occurs for an even number of electrons which has been the assumption of most studies to date~\cite{evenodd_peterson, pjds}. 

Although Pf and APf have identical non-Abelian braiding properties in the bulk, they represent distinct phases of matter as reflected e.g., in their edge physics~\cite{milica, wan} --- a signature of the underlying 
topological order~\cite{wen}.  A number of recent experiments have focused on measuring the quasiparticle charge at $\nu=5/2$~\cite{charge1,charge2,charge3}, and on detecting the non-Abelian statistics using edge-tunneling interferometry~\cite{statistics_willett,statistics_kang}.  These probes, while not definitive in all regards,  are consistent 
with the non-Abelian statistics, and
have provided additional insights into the nature of the ground state. In particular,
 the discovery of the counter-propagating mode~\cite{neutralmode} is consistent only with APf.

At the same time, other experiments have probed the stability of $\nu=5/2$ by driving the transitions from the incompressible to the compressible phases such as the Fermi liquid-like state~\cite{rr_cfl, HLR}, and the anisotropic, stripe and nematic phases~\cite{rh,xia}. This was accomplished by tilting the magnetic field~\cite{xia}, and by tuning the density in wide quantum well (WQW) samples~\cite{nuebler, liu}. In the latter case, it was recently noticed~\cite{liu} that the quantized $\nu=5/2$ state becomes stronger in the vicinity of a transition to the Fermi liquid-like phase.

In this Letter we introduce a new method to study the physics of the Moore-Read state and its P-H conjugate. We consider a compact torus geometry~\cite{yhl, duncanpbc} with a ``quartered" many-body Brillouin zone (BZ) for both even and odd number of electrons $N_e$. Regardless of the parity of $N_e$, for the Moore-Read 3-body Hamiltonian we obtain a zero-energy and zero-momentum ground-state, with bosonic (``magneto-roton") and fermionic (``neutral fermion"~\cite{nf}) collective modes at \emph{fixed} $N_e$. This is in stark contrast e.g. with the spherical geometry~\cite{HaldaneSphere}, where the zero-energy ground-state only exists for even $N_e$, and the fermionic mode can only be obtained for odd $N_e$~\cite{nf}. The essential physical similarity of the even and odd $N_e$ cases on the torus enables us to restrict to the latter case when Pf and APf can be further classified by their invariance under discrete rotations in high symmetry Bravais lattices. For special odd values of $N_e$ we recover spontaneous P-H symmetry breaking in \emph{finite} systems as Pf and APf acquire different angular momenta compatible with periodic boundary conditions (PBCs). 
This formalism is then applied to a realistic model of a wide quantum well with two subbands, S1 (symmetric subband with $n=1$ LL form-factor) and A0 (antisymmetric subband with $n=0$ LL form-factor), where P-H symmetry is broken explicitly by the mixing of electronic subbands/LLs as a result of tuning the density~\cite{liu}. We identify APf as the one describing the ground state in these circumstances, and show that its gap increases prior to the transition to the compressible phase, in agreement with experiments~\cite{liu}.

We consider $N_e$ electrons in a fundamental domain $\mathbf{L}_1\times \mathbf{L}_2$ subject to magnetic field $B\hat{z}$. 
An operator that translates a single electron and commutes with the Hamiltonian is the magnetic translation operator which obeys a non-commutative algebra, leading to the quantization of the flux $N_\Phi$ threading the system, $\hat{z}\cdot(\mathbf{L}_1 \times \mathbf{L}_2)=2\pi \ell_B^2 N_\Phi$, where $\ell_B=\sqrt{\hbar/eB}$ is the magnetic length. In a many-body system, symmetry classification is achieved by the help of the emergent many-body translation operators~\cite{duncanpbc}, which can be factorized into a center of mass and a relative part. The action of the former produces a characteristic degeneracy equal to $q$, where $N_e=pN$, $N_\Phi=qN$, and $N$ is canonically assumed to be the greatest common divisor of $N_e$ and $N_\Phi$ such that $p$ and $q$ are coprime. The eigenvalue of the relative translation operator is a many-body momentum $\mathbf{k}$~\cite{duncanpbc} that fully classifies the spectrum in the BZ $N \times N$, with the exception of high symmetry points where discrete symmetries may produce additional degeneracies, as we explain below.  

\begin{figure}[htb]
\includegraphics[scale=0.4]{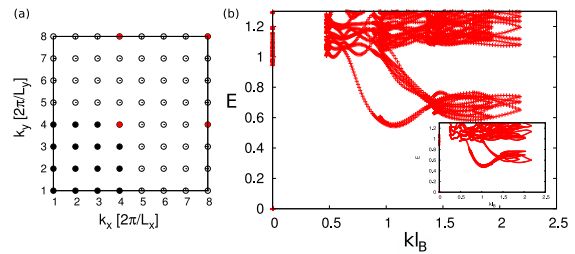}
\vspace{-1.5mm}
\caption[]{(Color online) (a) An example of a squared many-body BZ (open circles), and the folding to a ``quartered" BZ (black circles) for even $N_e$. Red circles denote sectors that map to the zero momentum $\mathbf{k}=0$. (b) Energy spectrum of the Moore-Read 3-body Hamiltonian in a quartered BZ for $N_e=14$ and $N_e=13$ (inset) particles.}
\label{fig:pf}
\vspace{-0pt}
\end{figure}

For the outlined algebraic derivation it is not essential that $p,q$ be coprime numbers. In fact, enforcing this condition might hide the important physical features of a FQH state. This occurs for the Moore-Read state which possesses a pairing structure revealed in its fundamental root pattern $110011001100\ldots$, that defines the clustering properties on genus-0 surfaces~\cite{jack}. The corresponding filling factor $\nu=1/2$ should be viewed as $\nu=2/4$ because the root pattern admits two particles in each four consecutive orbitals. To incorporate this clustering condition, we need to map the wavevectors onto a ``quartered" BZ $\tilde{N}\times \tilde{N}$~\cite{duncanunpublished}, which can be viewed as a result of ``folding" the original $N\times N$ zone, Fig.~\ref{fig:pf}(a). For $N$ even, the zone corner and midpoints of the zone sides all map to $\mathbf{k}=0$ point (red points in Fig.~\ref{fig:pf}(a)). These wavevectors also define the sectors of the Hilbert space where a three-fold degenerate Moore-Read groundstate is obtained~\cite{mr}. Moore-Read parent Hamiltonian also possesses a zero-energy ground state for an odd number of electrons, which corresponds to a single unpaired electron with $\mathbf{k}=0$~\cite{rg}. In this case, the folding still ``compactifies" the original BZ, but the $\mathbf{k}=0$ sector remains invariant under the folding.
Thus, in the quartered BZ, the ground-states of the Moore-Read Hamiltonian are invariably obtained in the $\mathbf{k}=0$ sector of the Hilbert space, as they should be for an incompressible liquid, and the 3-fold degenerate states are allowed to mix.

Advantages of the quartered BZ become obvious when the full energy spectrum of the Moore-Read 3-body Hamiltonian 
$
H_{\rm 3b} = -\sum_{i<j<k}\mathcal{S}_{ijk}\left[\nabla_i^4 \nabla_j^2 \delta(\mathbf{r}_i-\mathbf{r}_j)\delta(\mathbf{r}_j-\mathbf{r}_k) \right]
$
is studied as a function of momentum, Fig.~\ref{fig:pf}(b). Using the conventional definition of the BZ, there is no obvious structure in the low-lying excitation spectrum of the Moore-Read 3-body Hamiltonian. However, if the same spectrum is replotted in a quartered zone, 
it reveals a bosonic mode (the ``magneto-roton") and a fermionic mode (``neutral fermion")~\cite{nf}. Because of the BZ folding, in the even case ($N_e=14$) we obtain three copies of the bosonic mode and a single copy of the fermionic one. In the odd case ($N_e=13$), the multiplicities are interchanged~\cite{duncanunpublished}. 
Contrary to spherical geometry where spectra for different particle numbers have to be superimposed on the same plot to resolve the two modes, on the torus both modes are obtained for a fixed system size $N_e$. Additional advantage of PBCs is that one can access a quasi-continuum of the momenta $\mathbf{k}$, as opposed to a much smaller subset of angular momenta on the sphere. This is achieved by adiabatic variation of the shape of the unit cell in terms of its aspect ratio $|\mathbf{L}_1|/|\mathbf{L}_2|$ or the angle between vectors $\mathbf{L}_1$ and $\mathbf{L}_2$, subject to a constraint that the area $|\mathbf{L}_1 \times \mathbf{L}_2|$ remains fixed and equal to $2\pi\ell_B^2 N_\Phi$. In Fig.~\ref{fig:pf} we set the ratio equal to unity, and vary the angle between the square and the hexagon.   

\begin{figure}[htb]
\includegraphics[angle=0,scale=0.62]{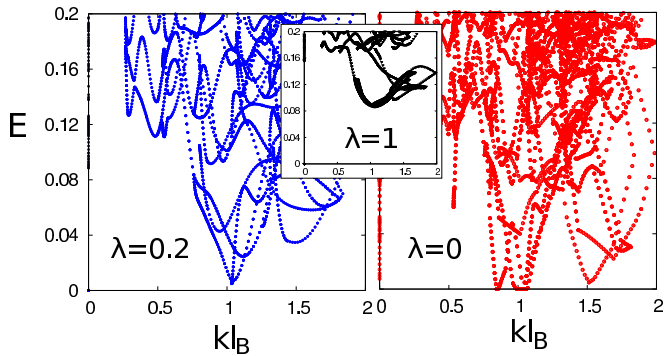}
\vspace{-2.5mm}
\caption[]{(Color online) Transition between the Moore-Read and the CFL state. Energy spectrum of $\lambda H_{\rm 3b}+(1-\lambda)H_{\rm C}$ is plotted as a function of momentum for $\lambda=0.2$ (left) and $\lambda=0$ (right), illustrating the collapse of the neutral mode around $k\sim \ell_B^{-1}$. Inset shows the spectrum for pure $H_{\rm 3b}$.}
\label{fig:pfcfl}
\end{figure}

Using the quartered BZ we can also directly address the phase transition between the Moore-Read state and the composite Fermi liquid (CFL). Spherical geometry is inadequate for this purpose because the two states have different ``shifts"~\cite{shift}. To capture the transition, we study the energy spectrum of a Hamiltonian that interpolates between the model 3-body Hamiltonian and the $n=0$ LL Coulomb interaction, $\lambda H_{\rm 3b}+(1-\lambda)H_{\rm C}$. In a quartered BZ, the spectra for even and odd $N_e$ are strikingly similar, and data in Fig.~\ref{fig:pfcfl} corresponds to $N_e=11$ which is also studied in a different model below. The neutral fermion and the magneto-roton modes become significantly distorted and difficult to identify for $\lambda<1$, nevertheless one can track their evolution until the eventual collapse of the gap for $\lambda \to 0$. At this point the ground state moves from $k=0$ to some $k \sim \ell_B^{-1}$, and the system undergoes a second-order transition to the compressible phase. In contrast, for $n=1$ LL Coulomb interaction, similar calculation does not lead to the gap closing for any $\lambda$. Adiabatic variation of PBCs also enables one to calculate e.g. the Hall viscosity, which is expected to diverge at $\lambda=0$ point. Regime of small $\lambda$ might display an interesting crossover between type-I and type-II superconducting behavior~\cite{sondhi}.    

In order to address the competition between Pf and APf for the generic (2-body) interactions, 
it is essential to consider all possible symmetries of the Hamiltonian at half filling. Particularly, 
the two symmetries of interest here are the P-H conjugation $\tau_{ph}$, and discrete rotations compatible
with PBCs~\cite{papicetal, yangle}.  The first is represented by an 
anti-unitary operator formally similar to the well-known case of time-reversal 
operator~\cite{tinkham_GT}. In the presence of both symmetries, extra (isolated) degeracies may occur 
which belong to conjugate or co-representations of discrete rotations~\cite{tinkham_GT, duncanpbc}.
In the case of even $N_e$, rotational symmetry does not lead to any additional quantum numbers for the 
ground state at $\nu=5/2$. For some ${\bf k}=0$ excited states, $\tau_{ph}^2=-1$ and these, following Kramer's 
theorem, are all
doubly degenerate. On the other hand, for odd $N_e$ and ${\bf k=}0$, $\tau_{ph}^2=1$ 
and only isolated degeneracies are
possible.  The ground state in this case is either unique or a doublet.
For any geometry other than hexagonal, the ground 
state is a singlet.  For hexagonal geometry, the ground state is also a singlet if the number 
of electrons is given by $N_e=6m+1$, where $m\in \mathbb{Z}$.  For other $N_e$, the ground state is a doublet~\cite{foot1}.

To understand these trends for generic Hamiltonians
it is helpful to consider the rotational properties of Pf and APf model states. On the torus with $n$-fold point symmetry, 
the
angular momentum of APf, measured relative to Pf, is given by $\Delta M=2N_{pair} ({\rm mod}\;n)$, 
where $N_{pair}$ is the number of paired electrons. 
For $n=2$ or $n=4$ (square or lower symmetry), $\Delta M=0$
since $N_{pair}$ is always even.  Thus there is no symmetry reason
for Pf and APf (being the eigenstates of different Hamiltonians) to be orthogonal.  On the other hand, for 
$n=6$, $\Delta M\not=0$ if $N_e\not=6m+1$, and the two states are necessarily orthogonal.  It is under precisely 
these conditions that the doublet ground states are observed.  These seemingly unrelated events are another
confirmation that in the case of the  Coulomb interactions the system is in the Moore-Read
phase. There are always doublets present in the spectrum, but they describe the ground state only
in cases where Pf and APf have $\Delta M\not=0$. Evidently, the rotational quantum numbers of the doublet
match those of Pf and APf. There will be no spontaneous breaking of the P-H symmetry in the absence of
such degeneracies until the thermodynamic limit is reached.  Achieving this property 
for finite sizes makes the comparison of the exact ground state with Pf or APf much cleaner than 
the case of even $N_e$, for example. As we show below, LL mixing splits 
the doublets in such a way that each member has a finite overlap with either Pf or APf, 
while having zero overlap with the other. 

\begin{figure*}[t]
  \begin{minipage}[l]{\linewidth}
    \includegraphics[scale=0.45]{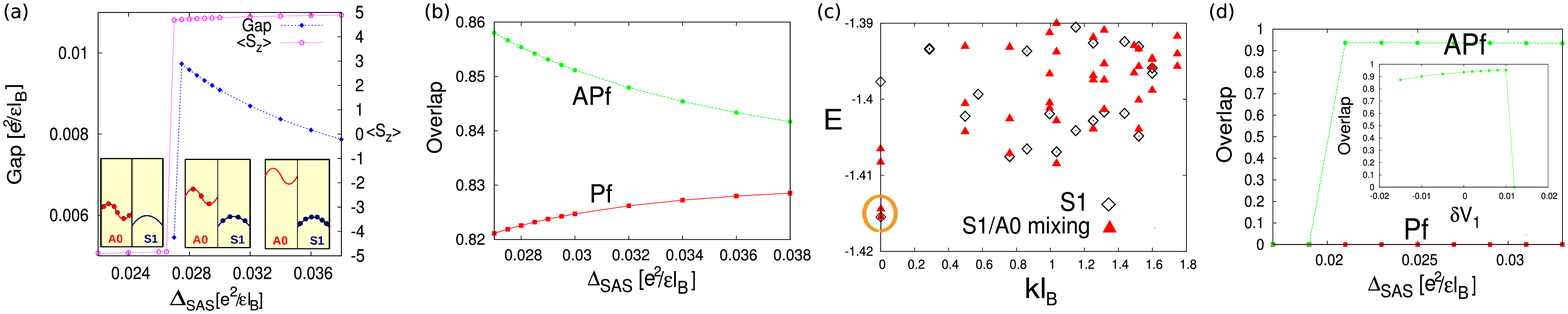}
  \end{minipage}
\caption{(Color online) (a) Neutral gap and mean value of the pseudospin $S_z$ for $N_e=10$ particles at half filling in a WQW ($w/\ell_B=2.7$). Transition is driven by changing $\Delta_{\rm SAS}$, and relative position of the electronic subbands is schematically shown in the inset. (b) Overlap between the exact $N_e=10$ ground state (projected onto S1 subband) and the Pf/APf wavefunctions. In the transition region ($\Delta_{\rm SAS}\approx0.028$) APf becomes favored. (c) Energy spectrum for infinite $\Delta_{SAS}$ (black diamonds) with an exact doublet at $k=0$, and the splitting of the doublet when $\Delta_{\rm SAS}$ is reduced (red triangles). (d) Same as in (b) but for $N_e=11$, when Pf and APf are mutually orthogonal. $\Delta_{\rm SAS}$ breaks PH symmetry and unambiguously selects APf (95\% overlap) over Pf (zero overlap). Inset shows the effect of varying the $V_1$ pseudopotential.}
\label{fig:wqw}
\vspace{-0pt}
\end{figure*}
We illustrate the ideas above on a model of the WQW in which electrons can populate two ``active" subbands with LL indices $n=0$ and $n=1$~\cite{papic_wqw}. We refer to these levels as A0 and S1, where S,A stands for the wavefunction in the perpendicular $z \in \left[0,w \right]$ direction, $w$ being the width of the well. For simplicity, we assume reflection symmetry around $z=w/2$, and the two subbands are given by symmetric/antisymmetric infinite square well wavefunctions $\phi_{\rm S}=\sqrt{2/w}\sin (\pi z/w)$, $\phi_{\rm A}=\sqrt{2/w}\sin (2\pi z/w)$. The remaining subbands are either completely filled or completely empty, and excitations to them are forbidden. Energy splitting between the subbands, $\Delta_{\rm SAS}$, can be realistically tuned by electrostatic gates~\cite{liu}, but in our calculation it is assumed to be an independent parameter. This model is expected to provide a realistic description of a number of recent experiments on the WQWs where FQH states were probed by tuning the effective interaction via subband/LL mixing. In particular, we focus on $\nu=5/2$ considering a half-filled S1 level mixing with an A0 level. It is assumed that the electron spin is fully polarized~\cite{morf, feiguin, spin1, spin2, spin3}.

In Fig.~\ref{fig:wqw}(a) we plot the neutral gap and mean value of the pseudospin operator $S_z=\frac{1}{2}\sum_i (c_{{\rm S},i}^\dagger c_{{\rm S},i} - c_{{\rm A},i}^\dagger c_{{\rm A},i})$ in the ground state for $N_e=10$ particles. We set $w/\ell_B=2.7$, which roughly agrees with the experimental width~\cite{xia, liu}. When $\Delta_{SAS}$ is large, excitations to A0 level are costly, and the ground state is fully polarized in S1 level and it is of the same nature as the one that is usually observed in wide samples. As $\Delta_{\rm SAS}$ becomes smaller, the difference of $n=0$ and $n=1$ LL form factors makes it increasibly favorable to promote particles into A0 level and reduce the correlation energy. Eventually, all particles migrate to A0 subband, where they form a composite Fermi liquid; as shown in Fig.~\ref{fig:wqw}(a), this happens slightly before the actual coincidence of the two subbands. The step-like behavior of $\langle S_z \rangle$ suggests the transition to be very sharp, and we find it to be more affected by the difference in LL (rather than the subband) form-factors. Right before the transition, excitations from S1 to A0 lead to an increase of the neutral gap of the system~\cite{liu}. We believe the increase of the gap to be an intrinsic feature of the system, but limitations on the system sizes attainable by exact diagonalization prevent us from performing proper finite-size scaling of the gap. 

Finally, we investigate the nature of the ground state before the transition to the compressible phase. In Fig.~\ref{fig:wqw}(b) we compare the $N_e=10$ electron ground state (projected to S1 level) with the Pf and APf wavefunctions defined on a hexagonal unit cell. At large $\Delta_{\rm SAS}$, the PH symmetry is preserved, and Pf/APf have identical overlaps with the exact state. In the transition region, PH symmetry is lifted, and APf overlap is growing while that of the Pf is decreasing. However, Pf and APf have a signicant overlap with each other for $N_e=10$ particles~\cite{llmixing2}. To avoid this problem, we instead consider $N_e=11$ case, when discrete symmetry forces Pf and APf to be orthogonal. At the same time, the Coulomb ground state, which is exactly twofold degenerate for infinite $\Delta_{\rm SAS}$, splits when $\Delta_{\rm SAS}$ is reduced (Fig.~\ref{fig:wqw}(c)). The size of the full Hilbert space for $N_e=11$ is very large to be directly diagonalized, so we limit the number of excitations to the higher subband~\cite{llmixing2}. The convergence is found to be rapid, and the essential properties of the ground state and the transition point can be captured accurately by allowing for not more than 3 electrons in a higher subband. The effect of a small amount of symmetry-breaking by $\Delta_{\rm SAS}$ is sufficient to select APf with 95\% overlap with the ground state, while the Pf overlap drops to zero (Fig.~\ref{fig:wqw}(d)). The APf overlap can be increased further by varying the $V_1$ pseudopotential in a ``SU(2)-invariant" manner~\cite{llmixing2} (Fig.~\ref{fig:wqw}(d), inset). These results provide unambiguous evidence that the ground state of the WQW is described by APf. 
    
In summary, we have demonstrated that PBCs for special finite systems with an appropriately defined BZ are essential for the study of the non-Abelian FQH states, their collective excitation spectra, and the effects of particle-hole symmetry breaking in WQWs under LL/subband mixing. Although we have presented results for the case of S1/A0 mixing, the more common case of S1/S2 mixing (applicable to narrower quantum well samples) similarly selects APf to describe the ground-state. This conclusion is not affected by mixing in even higher LLs (S3 and beyond). The sharpness of the transition and the ``population inversion" illustrated in Fig.~\ref{fig:wqw} might be relevant for the emergence of lowest LL physics under extreme tilting of the magnetic field~\cite{xia}. 

{\sl Acknowledgements}. We thank J. Eisenstein, M. Levin, N. Read, S. H. Simon,
and in particular Y. Liu and M. Shayegan for valuable
discussions. Z. P. would like to thank A. Sterdyniak,
N. Regnault, and B. A. Bernevig for useful comments. We
acknowledge support by DOE Grant No. DE-SC$0002140$,
and the Keck Foundation.

\bibliography{paper}

\begin{thebibliography}{99}
\bibitem{foot1}
In some cases the doublet is not the ground state but very near it.  Tweaking the interaction
by varying the thickness of the 2D layer or varying the pseudo-potentials makes it the ground
state.
\end{thebibliography}

\end{document}